\documentclass[twocolumn,showpacs,preprintnumbers,amsmath,amssymb,superscriptaddress]{revtex4}

\usepackage{amsmath,amsfonts,amssymb}
\usepackage[english]{babel} 
\usepackage[latin1]{inputenc} 
\usepackage[T1]{fontenc}
\usepackage{color}
\usepackage{float}
\usepackage{verbatim}
\usepackage{graphicx}
\usepackage{bm}
\usepackage{mathtools}
\usepackage{stmaryrd} 
\usepackage{anyfontsize}

\def \equi#1{\mathrel{\mathop{\kern 0pt\sim}\limits_{#1}}}

\begin{document}

\author{M. Chupeau}
\affiliation{Laboratoire de Physique Th\'eorique de la Mati\`ere Condens\'ee (UMR CNRS 7600), 
Universit\'e Pierre et Marie Curie, 4 Place Jussieu, 75255 Paris Cedex France}

\author{O. B\'enichou}
\affiliation{Laboratoire de Physique Th\'eorique de la Mati\`ere 
Condens\'ee (UMR CNRS 7600), Universit\'e Pierre et Marie Curie, 4 
Place Jussieu, 75255 Paris Cedex France}

\author{S. Redner} \affiliation{Santa Fe Institute, 1399 Hyde Park Road,
  Santa Fe, NM 87501, USA}

\title{Universality classes of foraging with resource renewal}

\begin{abstract}
  We determine the impact of resource renewal on the lifetime of a forager
  that depletes its environment and starves if it wanders too long without
  eating.  In the framework of a minimal starving random walk model with
  resource renewal, there are three universal classes of behavior as a
  function of the renewal time.  For sufficiently rapid renewal, foragers are
  immortal, while foragers have a finite lifetime otherwise.  In the specific
  case of one dimension, there is a third regime, for sufficiently slow
  renewal, in which the lifetime of the forager is independent of the renewal
  time.  We outline an enumeration method to determine the mean lifetime of
  the forager in the mortal regime.

\end{abstract}
\pacs{87.23.Cc, 05.40.Jc}

\maketitle

\section{Introduction}

What is the impact of renewal of resources on the state of a forager?  If the
environment is harsh and resources regenerate slowly, foragers may be
confronted by perpetual scarcity.  Thus a forager may often go hungry or even
starve.  Conversely, in an abundant environment where resources are quickly
replenished, a forager may never experience starvation risk.  Our goal is to
map out the states of a forager as a function of the renewal time within the
framework of the minimal ``starving random walk'' model~\cite{Benichou:2014}
that we define below.

The random walk model has been frequently invoked to describe the motion of
foraging animals~\cite{Berg:1983,Bartumeus:2005,Codling:2008}, as well as a
wide variety of classic applications~\cite{Weiss, Hughes, BenAvraham,
  Okubo,Bell:1970}.  In the wild, foraging animals can die from many causes,
such as diseases or old age~\cite{Fey:2015}.  The underlying dynamics from
these causes of death can be described by a ``mortal'' random walk that dies
according to a specified lifetime
distribution~\cite{Yuste:2006,Yuste:2013,Abad:2013}.  This model also a
variety of applications to diverse fields, such as the diffusion of light in
human tissue~\cite{Bonner:1987}, and biologically-inspired search
problems~\cite{Meerson:2015}.  In the context of foraging, an important
contributor to forager mortality is the possibility that it is unsuccessful
in its search for food within its habitat~\cite{Sinclair:1985}.  Thus the age
at which a forager dies is also determined by its trajectory and the amount
of available resources.  This coupling between the lifetime of a living
organism and its trajectory, along which environmental resources are
depleted, defines a nontrivial class of random-walk
problems~\cite{PW97,D99,BW03,ABV03,Z04,AR05}---including the starving random
walk model---that are relatively unexplored.

\begin{figure}
\includegraphics[width=0.475\textwidth]{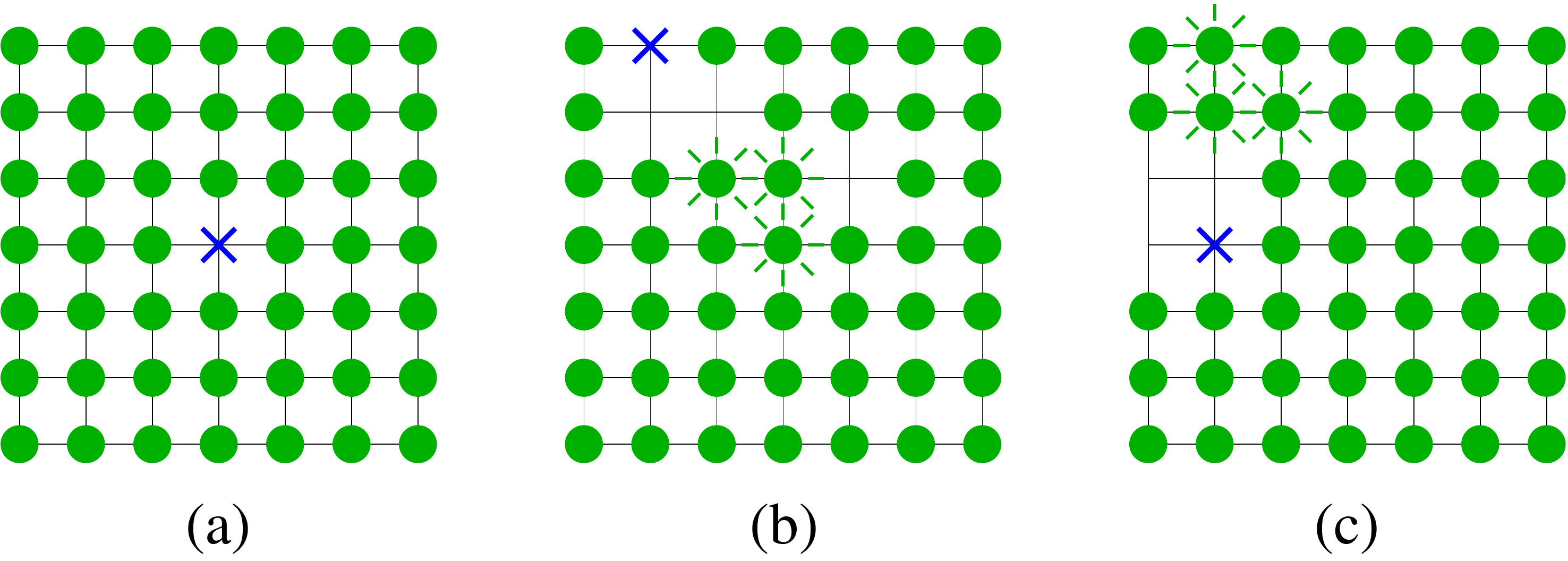}
\caption{Starving random walk with probabilistic resource renewal in two dimensions: (a)
  initial state and later times (b) \& (c).  Each site initially contains
  food that is eaten when found by the forager.  Food reappears on empty
  sites after a random renewal time (represented here as glowing circles).  The forager starves if it wanders
  $\mathcal{S}$ steps without eating.}
\label{model}
\end{figure}

In the original starving random walk model~\cite{Benichou:2014}, a random
walk irreversibly depletes its environment during its wanderings, and starves
if it wanders too long in a resource-depleted region.  Initially, each
lattice site contains one food unit.  Whenever a forager, which undergoes a
random walk, arrives at a full site, the food is completely eaten.  Once
depleted, a site remains empty.  Whenever the forager arrives at an empty
site, it does not eat.  If the forager goes $\mathcal{S}$ steps without
eating, it starves.  We can think of $\mathcal{S}$ as the metabolic capacity
of the forager---the amount of time it can live without food before starving.
Asymptotic expressions for the walker lifetime and the territory visited at
the starvation time were given in one dimension~\cite{Benichou:2014}.
Estimates for these two quantities in dimensions $d\geqslant 3$ and a lower
bound for the territory visited at starvation in $d=2$ were also given.
These results provide a first step in quantifying the interplay between the
trajectory of a forager and the consumption of food and their effect on the
lifetime of the forager.

However, the natural resources being consumed---preys, plants, water, and
nutrients---typically obey their own dynamics.  In particular, consumed
resources usually do not disappear
permanently~\cite{Evans:1972,Lemaire:2001}.  Instead, they typically
regenerate a certain time after they have been
depleted~\cite{McCallum:1905,Harrison:1980,Fenner}.  This basic fact
motivates our study of starving random walks with the possibility of renewal
of resources (Fig.~\ref{model}).

\begin{figure}
\resizebox{0.37\textwidth}{!}{\input{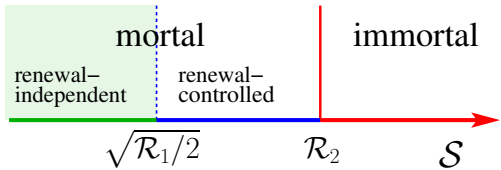}}
\caption{Model phase diagram.  The shaded zone, where the lifetime is
  independent of renewal rate, occurs only $d=1$; the two other regimes arise
  for any $d$.  The thresholds between these regimes are given by
  Eqs.~\eqref{bound} and \eqref{class3}.}
\label{diag_phase}
\end{figure}

  As we will discuss, resource renewal substantially modifies the properties
  of a starving random walk.  Using extreme trajectory arguments, we will
  argue that the correlations induced by the coupling between the trajectory
  of a forager, its metabolic capacity $\mathcal{S}$, and the dynamics of the
  resources lead to three universal regimes of behavior that are determined
  by $\mathcal{S}$ and the bounds $\mathcal{R}_1$ and $\mathcal{R}_2$ of the
  support of the renewal time distribution, but are insensitive to the shape
  of this distribution (Fig.~\ref{diag_phase}).  We will first demonstrate
  the existence of a transition between an immortal regime, in which the
  forager cannot starve, no matter what its trajectory, and a mortal regime,
  where the forager must eventually starve.  Both regimes arise in any
  spatial dimension.  We will also show that a third regime arises in one
  dimension only, in which renewal is so slow that the forager lifetime is
  the same as in the case of no renewal.  We also develop an enumeration
  method that yields, in principle, the exact value of the mean lifetime of a
  forager in the mortal regime.
  
We will first address the case of starving random walks with deterministic
renewal in one dimension (Sec.~II).  In Sec.~III, we extend our basic results
to the case of probabilistic renewal and to higher dimensions.  Some brief
conclusions are given in Sec.~IV.

\section{ Deterministic renewal in 1D}

Let us first investigate starving random walks with \emph{deterministic}
renewal in one dimension.  Such a deterministic mechanism describes plants
that grow at a fixed rate to reach an edible size a fixed time after having
been previously defoliated~\cite{Erickson:1976}.  We posit that food that is
eaten at time step $t$ reappears at time $t+\mathcal{R}$, with $\mathcal{R}$
an integer.  In each time step, the time elapsed since an empty site was
depleted is increased by one and food appears at any site where this time
equals $\mathcal{R}$.  As part of this same time step, the walker hops to one
of its nearest neighbors.  If the site contains food, which may have appeared
just before the walker arrives, the food is eaten.  We call the set of empty
sites, which may or may not be connected, the ``desert''.

\subsection{Immortality}
  
\begin{figure}
\centering
\includegraphics[width=0.45\textwidth]{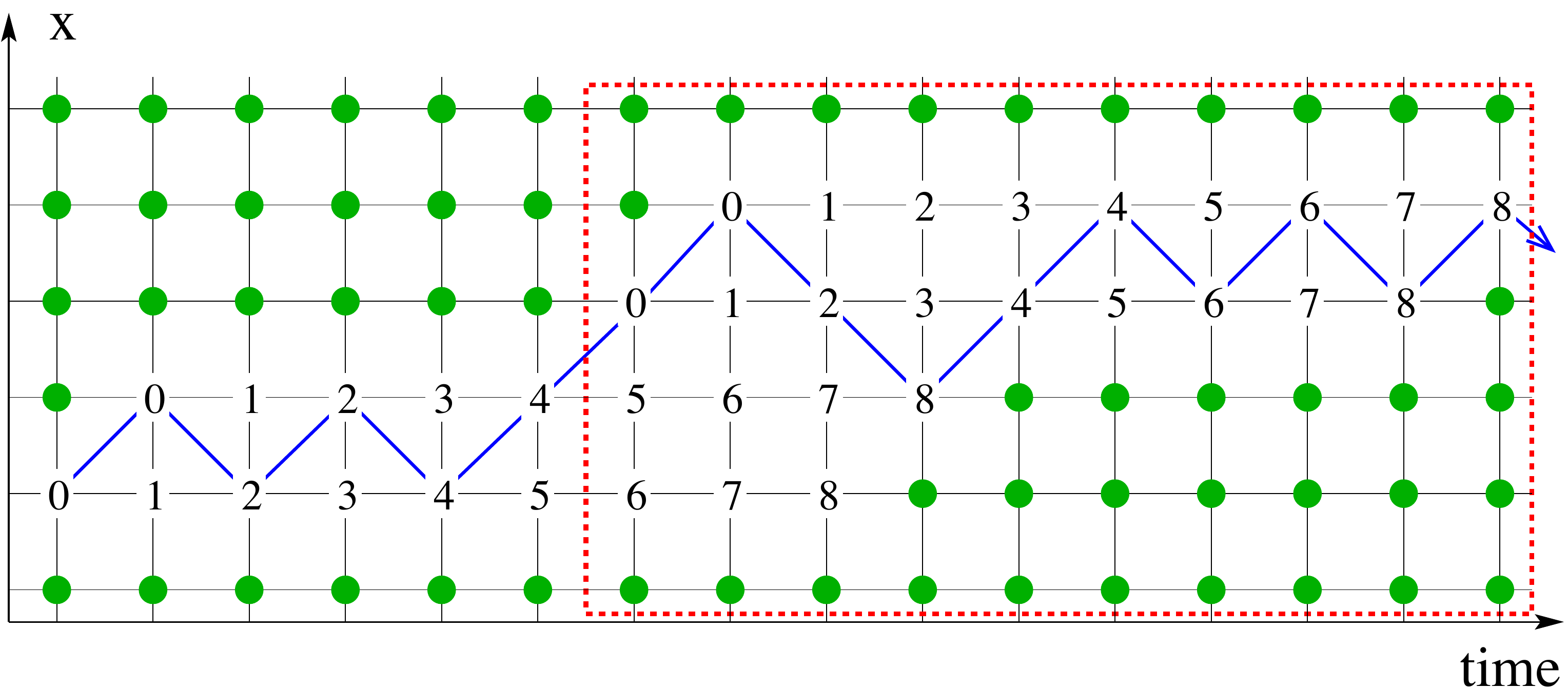}
\caption{Illustration of the common pattern (inside the dashed rectangle) of
  all extremal trajectories for the case of renewal time $\mathcal{R}=9$.
  This pattern consists of depleting two consecutive sites at one end of the
  desert.  The forager wanders inside this desert which gradually shortens
  until it reaches length 2. The pattern ends when the walker is certain to
  land on a food-containing site.  Green dots denote food-containing sites
  and the numbers give the ages of each site.}
\label{limit}
\end{figure}

For a forager with metabolic capacity $\mathcal{S} \in \mathbb{N}$, we now
determine the range of the renewal times for which immortality arises.  A
forager is immortal if it survives forever on any trajectory, and, in
particular, on the most unfavorable trajectories.  The set of such
trajectories is infinite, but they all possess the common pattern
(Fig.~\ref{limit}) that the walker depletes two sites in a row at one end of
the desert, before wandering within the desert as long as possible until
being certain, because of resource renewal, to land on a food-containing site
(see Appendix A for details).  It takes $\mathcal{R}$ time steps before food
reappears on one of these two sites (Fig.~\ref{limit}).  Thus, roughly
speaking, when $\mathcal{R}<\mathcal{S}$, the forager necessarily survives.
The precise criterion actually is $\mathcal{R}\leq \mathcal{R}^*$, with
\begin{equation}
\label{bound}
\mathcal{R}^* =
\begin{cases}
\mathcal{S}&\quad \mathcal{S} ~ \mathrm{even}\,,\\
\mathcal{S}+1&\quad \mathcal{S} ~ \mathrm{odd}\,,
\end{cases}
\end{equation}
due to the even-odd oscillations of a nearest-neighbor random walk (Appendix
A).

\subsection{Mean Lifetime in the Mortal Regime} 

On the other hand, when $\mathcal{R}> \mathcal{R}^*$, the forager is mortal
and thus eventually starves with probability 1 because we can construct
trajectories that necessarily lead to starvation.  These trajectories again
involve the forager first carving out a desert and then wandering strictly
within this desert so that renewal does not reach the forager before it
starves.  In contrast to the situation in Fig.~\ref{bound}, the walker
starves before it reaches a site where the resource has been regenerated.  By
their very existence, such trajectories are achieved with non-zero
probability.  From classical results about Markov chains (see, e.g.,
\cite{Grinstead:2012}), every trajectory will eventually generate a
configuration for which the forager starves.  Thus there are two regimes of
behavior---immortality and mortality.  The boundaries between these regimes
depend only on the metabolic capacity $\mathcal{S}$ of the forager and the
renewal time $\mathcal{R}$.

Using the Markov chain formalism, we can, in principle, determine the mean
lifetime of the forager by enumerating the configurations of the system as
the forager wanders.  A configuration is defined as the location and age of
each empty site in the desert, the position of the forager in the desert, and
the number of time steps elapsed since the forager last ate.  Here the age of
an empty site is the time since the food was last consumed at this site.
Thus a newly empty site has age 0, while a site that will regenerate at the
next step has age $\mathcal{R}-1$.  Because the desert has a finite size, the
number of configurations is finite.  We can therefore write the transition
matrix that describes the evolution of the system at each step of the forager
and thereby extract its mean lifetime.

\begin{figure}
\centering
\includegraphics[width=230pt]{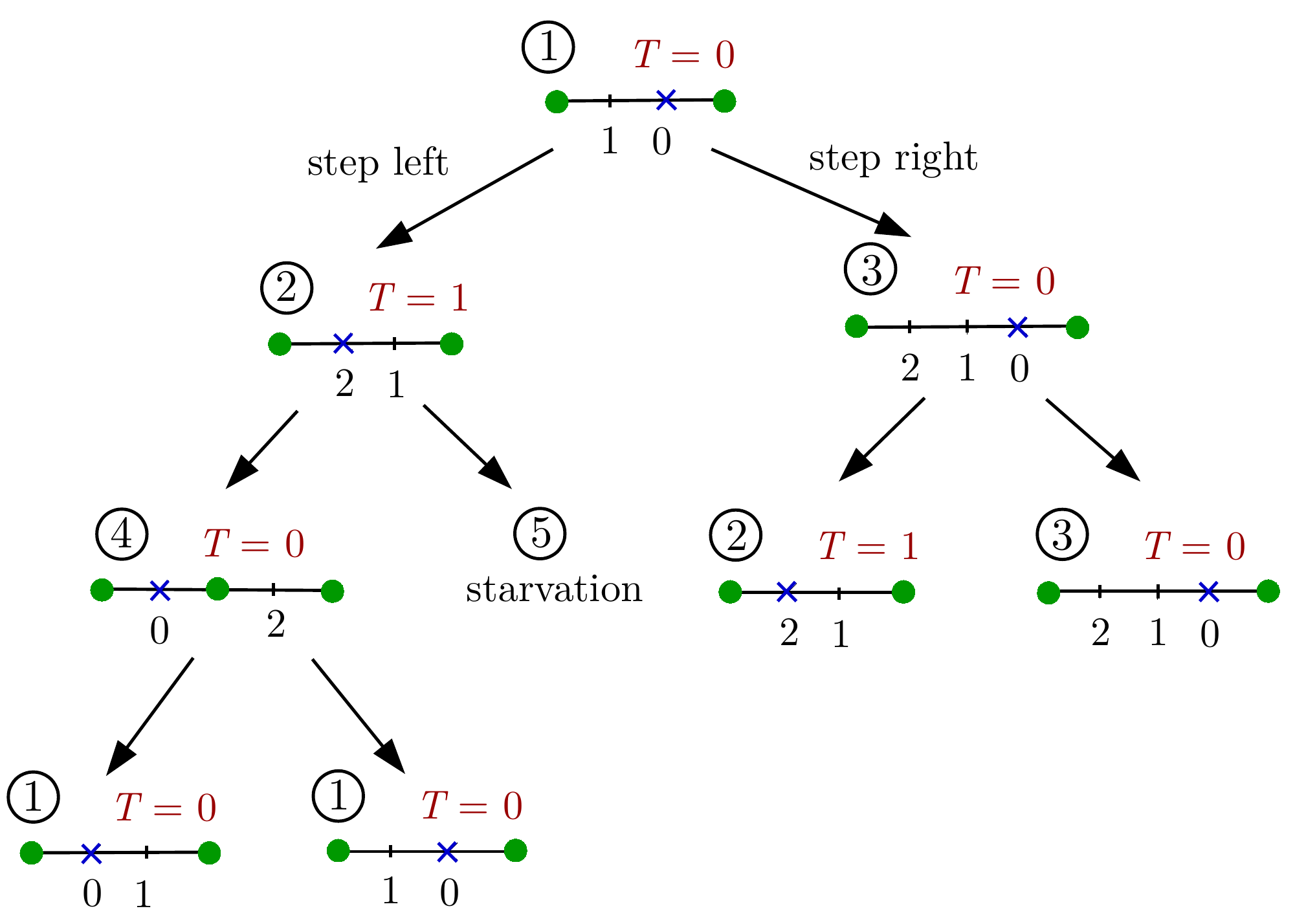}
\caption{State space of the system for metabolic capacity $\mathcal{S}=2$ and
  renewal time $\mathcal{R}=3$.  The circled numbers denote the
  distinct states and $T$ is the time elapsed since the forager last ate.}
\label{diagram_trans}
\end{figure}

Let us illustrate this enumeration for the simple case of $\mathcal{S}=2$ and
$\mathcal{R}=3$.  For this example, there are five distinct configurations
(Fig.~\ref{diagram_trans}).  State \textcircled{1} arises after the first
step, and the evolution of the system from one state to another is shown in
the figure. The associated transition matrix is
\begin{equation}
\begin{pmatrix}
 0 & 0 & 1 & 0 & 0 \\ 
 1/2 & 0 & 0 & 1/2 & 0 \\ 
 0 & 1/2 & 0 & 0 & 0 \\ 
 1/2 & 0 & 0 & 1/2 & 0 \\ 
 0 & 1/2 & 0 & 0 & 1
\end{pmatrix}
\equiv \begin{pmatrix}
 Q & 0 \\ 
 V & I
\end{pmatrix}
\end{equation}
where the states are listed in order \textcircled{1}--\textcircled{5}, with
\begin{equation}
 Q=\begin{pmatrix}
 0 & 0 & 1 & 0  \\ 
 1/2 & 0 & 0 & 1/2  \\ 
 0 & 1/2 & 0 & 0  \\ 
 1/2 & 0 & 0 & 1/2 
\end{pmatrix},
\end{equation}
and $V=(0,\frac{1}{2},0,0)$.  We define the matrix (see
\cite{Grinstead:2012})
\begin{equation}
 N\equiv (I- Q^+)^{-1} = \begin{pmatrix}
 2 & 2 & 1 & 2 \\ 
 1 & 2 & 1 & 1 \\ 
 2 & 2 & 2 & 2 \\ 
 1 & 2 & 1 & 3
\end{pmatrix},
\end{equation}
where $Q^+$ is the transpose of $Q$.  Each entry $N_{ij}$ in this matrix is
the average time that a system, which ultimately reaches starvation, spends
in configuration $j$ when it starts from configuration
$i$~\cite{Grinstead:2012}.  From this matrix we can extract the mean
absorption time $t_i$ starting from the state $i$.  These are given by
\begin{equation}
\begin{pmatrix}
 t_1 \\ 
 t_2 \\ 
 t_3 \\ 
 t_4
\end{pmatrix} = N \begin{pmatrix}
 1 \\ 
 1 \\ 
 1 \\ 
 1
\end{pmatrix} = \begin{pmatrix}
 7 \\ 
 5 \\ 
 8 \\ 
 7
\end{pmatrix}.
\end{equation}

Thus the mean lifetime of the forager is $t_1+1=8$, because after the first
step, the system is necessarily in the state $\textcircled{1}$.  In
principle, this method can be extended to higher dimensions and also to
probabilistic renewal with a bounded support of renewal times.  However, in
practice, this approach quickly becomes intractable because the number of
possible configurations becomes prohibitively large when both $\mathcal{R}$
and $\mathcal{S}$ are large.  Nevertheless, this approach gives a
well-defined prescription for computing the average time until the forager
starves.
 
\subsection{Renewal Independent Regime}  

In one dimension, the mortal regime class can be further divided in two
sub-regimes: (a) forager lifetime dependent on $\mathcal{R}$, and (b) renewal
independence---lifetime independent of $\mathcal{R}$.  Clearly, as
$\mathcal{R}$ increases, the forager lifetime decreases and approaches the
no-renewal limiting value as $\mathcal{R}\to\infty$.  Does this decrease stop
when $\mathcal{R}$ reaches a finite critical value $\mathcal{R}^{\dagger}$,
or does the decrease continue as $\mathcal{R}\to\infty$?  To resolve this
question, we determine if there is at least one trajectory for which the
forager can return to a replenished site without starving.  If there is such
a trajectory, then renewal is relevant, as the lifetime of the forager
depends on the renewal time.

For a forager to return to a site on which food is renewed requires: (i)
living long enough for such a renewal to occur and (ii) staying sufficiently
close to this site to reach it without starving.  These two attributes are
most easily satisfied for a site (which we take as the origin) at which food
has just been eaten and is surrounded by full sites.  We determine the
largest value of $\mathcal{R}$ for which the forager can return to this
origin after the food at this site has been renewed.

\begin{figure}
\centering
\includegraphics[width=160pt]{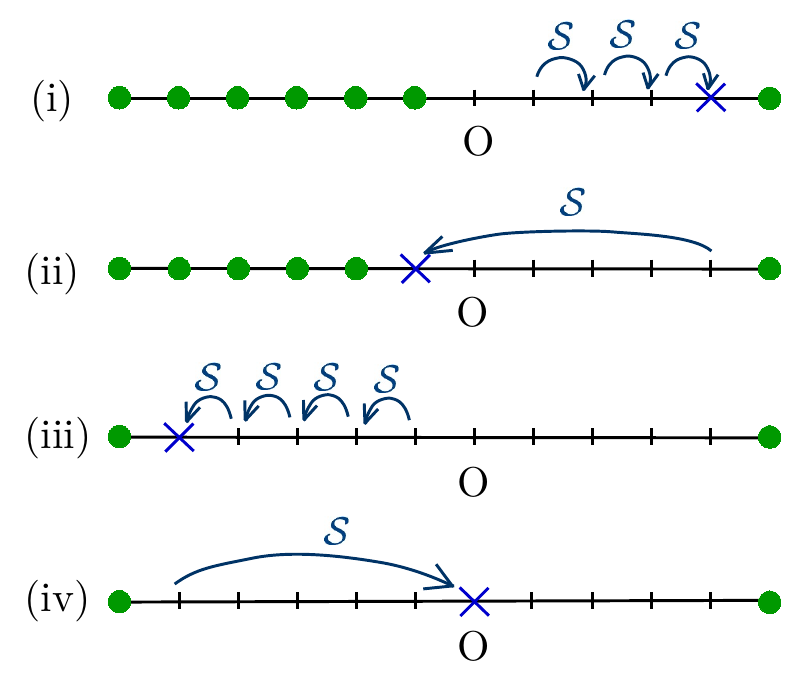}
\caption{Optimal trajectory for a forager to live the longest and still be
  able to return the origin O and consume the replenished resource on this
  site, for the case $\mathcal{S}=5$. (i) The forager eats a site on the
  right edge of the desert every $\mathcal{S}$ steps until carving the
  largest semi-desert that it can cross. (ii) The forager crosses the
  desert. (iii) Reflection of stage (i). (iv) The forager crosses the left
  semi-desert to reach the origin.  Such an excursion lasts
  $\mathcal{R}^{\dagger}$ steps (Eq.~\eqref{class3}). }
\label{optimal}
\end{figure} 
 
To return to the origin without starving imposes the constraint that the
forager does not stray more than $\mathcal{S}$ steps from the origin.
Moreover, to maximize the time that the forager wanders, it should eat only
when it really needs to, that is, every $\mathcal{S}$ steps.  The trajectory
on which the forager lives the longest while staying within $\mathcal{S}$
steps of the origin therefore consists of the following components
(Fig.~\ref{optimal}): (i)~The forager creates a semi-desert of length
$\mathcal{S}$ on one side, say to the right, of the origin.  During this
creation of the semi-desert, the forager eats by moving to a previously
unvisited site only every $\mathcal{S}$ steps.  (ii)~The forager successfully
crosses this semi-desert, which is the longest possible for which a
successful traversal is possible.  (iii)~The forager creates a mirror image
semi-desert of length $\mathcal{S}$ to the left of the origin.  (iv)~The
forager crosses this left semi-desert and fetches the regenerated food at the
origin.

The duration of this excursion is roughly $2 \mathcal{S}^2$, as the forager
has eaten $2 \mathcal{S}$ times when it returns to the origin.  The
enumeration of the above sequence of moves (Appendix B) gives the maximal
renewal time $\mathcal{R}^{\dagger}$ with
\begin{equation}\label{class3}
\mathcal{R}^{\dagger}= 
\begin{cases}
  2 \mathcal{S}^2-3\mathcal{S}+4 \qquad \textrm{$\mathcal{S}$ even}, \\
  2 \mathcal{S}^2-\mathcal{S}+1 \qquad~\; \textrm{$\mathcal{S}$ odd}.
\end{cases}
\end{equation}
For $\mathcal{R}\leq \mathcal{R}^\dagger$, a forager has a non-zero
probability to eat food at a site where renewal has occurred.  Conversely,
for $\mathcal{R}>\mathcal{R}^{\dagger}$, a forager cannot reach any
replenished site, so that renewal has no impact on the forager lifetime.
Hence the mean lifetime does not gradually converge to the limiting
no-renewal value, but rather reaches this value for
$\mathcal{R}=\mathcal{R}^{\dagger}+1$.  Thus we infer that there exists a
renewal-independent regime (Fig.~\ref{diag_phase}) for the lifetime of the
forager.

\section{Extensions}

We now extend the starving random walk model with resource renewal to
accommodate two ecologically realistic features: (i) probabilistic renewal,
in which the resource is regenerated at a random time after depletion, rather
than after a fixed time $\mathcal{R}$, and (ii) starving random walks with
resource renewal in higher dimensions.

\subsection{Probabilistic renewal}

Suppose that each empty site is replenished a time $\tau$ after the resource
at that site has been consumed, with $\tau$ drawn from a continuous
distribution with support
$[\mathcal{R}_1,\mathcal{R}_2]\subset \mathbb{R}^+$.  This means that for a
given site renewal cannot happen before a time $\mathcal{R}_1$ and also that
replenishment necessarily occurs within a time $\mathcal{R}_2$ after
depletion.  We make no assumption on the shape of this distribution.  In
particular, $\mathcal{R}_1$ can be zero and $\mathcal{R}_2$ can be infinite.

In the case of deterministic renewal (Sec.~II), we saw that the random walker
is immortal when it is certain to land on a food-containing site before
starving, even on the most unfavorable trajectories.  The criterion for
immortality in this case is determined by $\mathcal{R} \leq \mathcal{R}^*$,
with $\mathcal{R}^*$ given in Eq.~\eqref{bound}.  For probabilistic renewal,
the walker is sure to land on a food-containing site before starving if every
renewal time in the support of the renewal-time distribution is smaller than
$\mathcal{R}^*$.  Therefore, immortality is assured when
$\mathcal{R}_2 \leq \mathcal{R}^*$.  However, if the upper bound
$\mathcal{R}_2$ is infinite, that is, if replenishment on some sites can take
an arbitrarily long time, immortality cannot occur.

On the other hand, if $\mathcal{R}_2>\mathcal{R}^*$, there exist patterns of steps for the random walker that lead to starvation, as in the case of deterministic renewal. Hence the walker is mortal. Additionally, the enumeration method presented above for deterministic renewal
can be implement in a similar manner for the case of probabilistic renewal.
In the probabilistic case, however, food does not reappear at a fixed time
after depletion but at a time that is drawn from the renewal time
distribution.  What this means practically is that the number of
configurations in probabilistic renewal is larger than that for deterministic
renewal.  Moreover, if the support of the distribution of renewal times is
unbounded, the enumeration approach fails because the number of
configurations is infinite.

We also argued in Sec.~II that there exists a second transition inside the
mortal regime between a sub-regime in which dynamics of the renewal controls
the lifetime of the walker and a sub-regime where the lifetime becomes
independent of the renewal dynamics.  We inferred the existence of this
transition by constructing the extremal trajectory that demarcates this
second transition (Fig.~\ref{optimal}).  If renewal has not occurred at the
origin when the walker reaches this site at the end of the pattern of steps
of Fig.~\ref{optimal}, then renewal has no impact on the trajectory. Thus,
the renewal independent regime arises if every depleted site remains empty
for at least $\mathcal{R}^{\dagger}$ steps.  In the case of probabilistic
renewal, this second transition occurs when
{$\mathcal{R}_1 > \mathcal{R}^{\dagger}$}.

These results are summarized in the phase diagram of Fig.~\ref{diag_phase};
this is our key result.

\subsection{Higher dimensions}

We now turn to starving random walks on higher-dimensional lattices for the
general situation of probabilistic renewal.  The class of trajectories that
are the least favorable for the survival of walker (see Fig.~\ref{limit}),
still arises in higher dimensions.  Hence the immortality criterion
$\mathcal{R}_2 \leq \mathcal{R}^*$ derived in the previous subsection remains
valid, independent of the spatial dimension. Moreover, in the mortal regime,
the enumeration method still works and can be used to determine the mean
lifetime of a random walker.

In contrast to the transition to immortality, a transition to a
renewal-independent regime does not occur in higher dimensions.  The unique
feature of one dimension is that the walker \emph{must} traverse the desert
that was carved by its previous trajectory to reach replenished sites.  In
contrast, in higher dimensions, there always exist trajectories on which a
forager can stay alive for an arbitrarily long time and still return to the
replenished sites without starving, because it can avoid the desert instead
of having to cross it.  Thus the renewal time---no matter how long---always
affects the lifetime of the forager in greater than one dimension.  Thus in
higher dimensions there is only an immortal regime and a mortal regime in
which the lifetime is function of the renewal dynamics.  The transition
between these two regimes is determined only by the upper bound of the
distribution of renewal times, and not by the shape of this distribution, or
by the spatial dimension.

\section{Summary and conclusion}

To summarize, the renewal of resources has a dramatic effect on starving
random walks. There exist three regimes of behavior as a function of the
renewal time $\mathcal{R}$: (i) an immortal regime where a forager can live
forever, (ii) a mortal regime where the forager lifetime is finite and
depends on $\mathcal{R}$, and (iii) a renewal-independent mortal regime where
renewal does not affect the lifetime of a forager.  The latter arises only in
one dimension, in which the average forager lifetime equals the value
obtained in the absence of any renewal.  In contrast, regimes (i) and (ii)
arise in any spatial dimension and are universal with respect to the
distribution of renewal times.  The transitions between these regimes depend
only on the bounds of the support of the renewal-time distribution and not on
its shape.  Much of this new phenomenology is controlled by the \emph{times
  between visits} to distinct sites in a random walk, an apparently
unexplored feature of site visitation statistics of random walks.  Finally,
we outlined an enumeration method to determine the mean forager lifetime in
the mortal regimes (ii) and (iii).  Average values of other basic
observables, such as the number of distinct sites visited at starvation, the
number of sites in the desert, the time spent in a certain configuration, can
also be extracted from this approach.

Immortality is the main new feature that arises as a result of resource
renewal.  If a wandering organism can survive without food longer than the
time needed for resources to be replenished, its lifetime is no longer
limited by starvation but rather by external constraints (such as predation,
diseases, life expectancy of the species, etc.).  We speculate that perhaps
the metabolic capacity of a given species is determined by the characteristic
time for renewal of resources.

This work represents a first step to provide insight of the impact of
resource renewal on the fate of a forager that depletes its environment by
consumption.  While most of our qualitative analysis was specific to the case
of one dimension, our approach applies for any spatial dimension and also to
arbitrary renewal dynamics.  A basic question that we have not fully
addressed is the analytic determination of the mean lifetime of the walker in
the mortal regime.  This calculation is of particular importance in two
dimensions where it should be directly applicable to the modeling of animal
behavior.  In addition to developing a more complete theory in two
dimensions, the inclusion of additional realistic features to this model,
such as sensory awareness of the forager, or interactions between several
foragers, such as sharing resources, are needed to give a more realistic
description of ecosystems.
 
We acknowledge NSF Grant No. DMR-1205797 (S.~R.) and ERC starting Grant
No.\ FPTOpt-277998 (O.~B.) for partial support of this research. 


%

\newpage\pagebreak\newpage

\appendix

\section{The bound $\mathcal{R}^*$}

\begin{figure}[h]
\centering
\includegraphics[width=240pt]{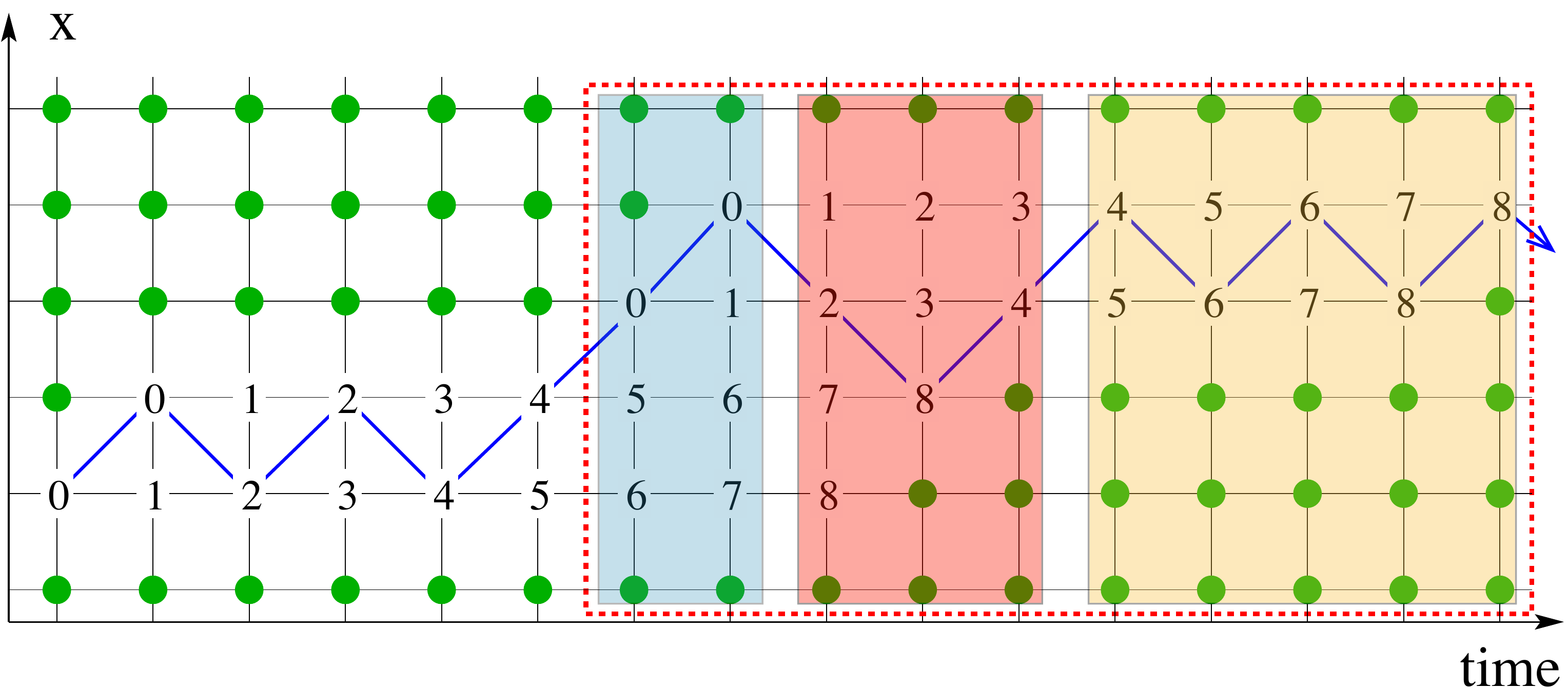}
\caption{More detail of the common pattern (inside the dashed rectangle) of
  all extremal trajectories for renewal time $\mathcal{R}=9$. This pattern
  starts by depleting two consecutive sites at one end of the desert (shaded
  in blue). The forager then wanders inside this desert which gradually
  shortens (shaded in red) until it reaches length 2 (shaded in yellow). The
  pattern ends when the walker is sure to land on a food-containing
  site. Green dots denote food-containing sites and the numbers give the ages
  of each site.}
\label{pattern}
\end{figure}

We provide the details for the determination of Eq.~\eqref{bound}, in the
case of deterministic renewal.  This bound for $\mathcal{R}^*$ depends on the
parity of the metabolic capacity $\mathcal{S}$ because of the even-odd
oscillations of the nearest-neighbor random walk.  As described in the main
text, the most unfavorable trajectories---on which the walker remains the
longest without eating---possess the common pattern of eating two consecutive
sites at one edge of the desert and then wandering as long as possible inside
the desert, without depleting any additional site.  This last feature implies
that the desert gradually shortens as renewal happens, finally confining
these most unfavorable trajectories to a two-site desert, made of the first
two depleted sites of the pattern (see Fig.~\ref{pattern}).

\begin{figure}[h]
\centering
\includegraphics[width=250pt]{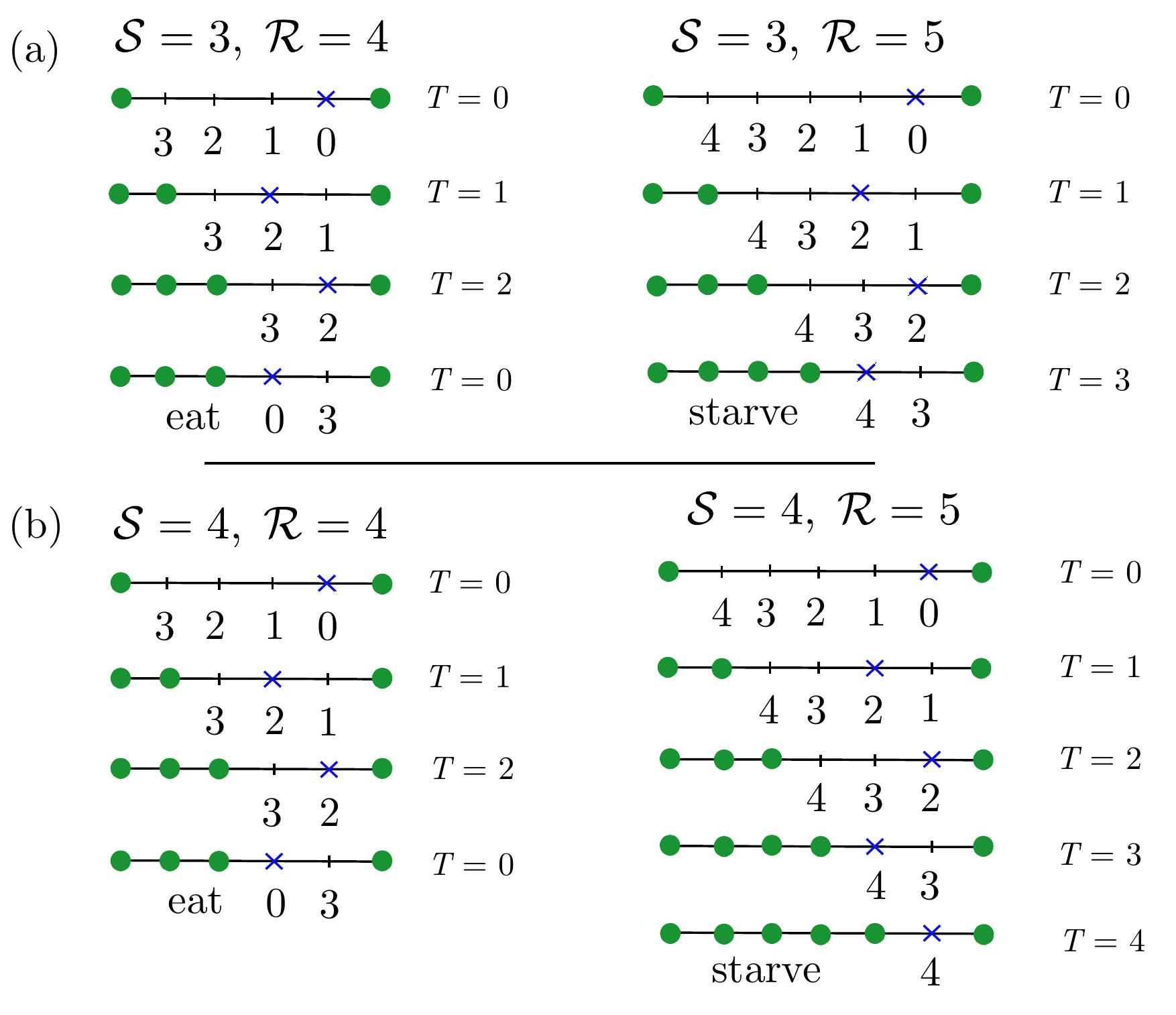}
\caption{Destiny of the walker on the most unfavorable trajectories, for:
  (left column) the maximal renewal time $\mathcal{R}^*$ that leads to
  immortality, and (right column) for the minimal value of $\mathcal{R}$ that
  leads to mortality.  The cases of odd and even metabolic capacities
  $\mathcal{S}$ are shown in (a) and (b), respectively.  For both even and
  odd $\mathcal{S}$, the maximal value of $\mathcal{R}$ that yields
  immortality is the smallest even integer equal to or greater than
  $\mathcal{S}$.  Here $T$ denotes the number of steps since the last meal.
  The dots represent food and the cross indicates the walker.  Empty sites
  are labeled by the time elapsed since the food on these sites was eaten.}
\label{threshold1}
\end{figure}

Depending on the value of $\mathcal{S}$, the walker either survives long
enough to eat when renewal happens on these two sites (in the immortal
regime), or starves (in the mortal regime).  We determine the maximal value
of the renewal time $\mathcal{R}^*$ that corresponds to the immortal regime
for an example of most unfavorable trajectory for two consecutive values of
$\mathcal{S}$ (Fig.~\ref{threshold1}).  The walker is sure to eat before
starving when the renewal time is $\mathcal{R}=\mathcal{R}^*$, even on this
unfavorable trajectory (left column of Fig.~\ref{threshold1}), but can die if
$\mathcal{R}=\mathcal{R}^*+1$ (right column of Fig.~\ref{threshold1}).  For
this example, which can be generalized to every value of $\mathcal{S}$, we
see that
\begin{equation}
\label{bound1}
\mathcal{R}^* =
\begin{cases}
\mathcal{S}&\qquad \mathcal{S} ~ \mathrm{even}\,,\\
\mathcal{S}+1&\qquad \mathcal{S} ~ \mathrm{odd}\,.
\end{cases}
\end{equation}

\section{The bound $\mathcal{R}^{\dagger}$}

We now give the details to derive Eq.~(6) for the case of deterministic
renewal.  Note that the bound for $\mathcal{R}^{\dagger}$ also depends on the
parity of the metabolic capacity $\mathcal{S}$ because of the even-odd
oscillations of the nearest-neighbor random walk. If renewal is sufficiently
quick, there exist trajectories for which the walker can return to a
replenished site, in particular the origin of the walk, before starving.  On
the other hand, if renewal is too slow, then the walker either dies before
this renewal happens, or carves a desert that is too large to be crossed
without starving.

As mentioned in the main text, the maximal value $\mathcal{R}^{\dagger}$ of
the renewal time for which the walker has a chance to return to the origin
after the resource on this site has been renewed requires that: (i) the
walker lives long enough for this renewal to occur, and (ii) the walker must
stay sufficiently close to the origin to be able to reach it without
starving.  The walker can satisfy these two constraints by eating
approximately every $\mathcal{S}$ steps and by staying within a segment of
size $2 \mathcal{S}$ centered on the origin.

The trajectories on which the walker lives the longest while remaining within
$\mathcal{S}$ steps of the origin consist of the following
(Fig.~\ref{limit}): (i) The walker creates a desert of $\mathcal{S}$ sites on
one side of the origin, say the right.  During this creation of the desert,
the walker waits as long as possible between each meal, that is to say
$\mathcal{S}$ steps if $\mathcal{S}$ is odd, or $\mathcal{S}-1$ steps if
$\mathcal{S}$ is even.  Indeed, starting from the right edge of the desert,
the walker needs an even number of steps to come back to this edge; thus an
odd number of steps is required to eat (and deplete) the resource at the next
site on the right side.  (ii) The walker crosses the desert and reaches the
site to the left of the origin after $\mathcal{S}$ steps. (iii) The walker
creates a mirror image desert on the left side of the origin, by depleting
$\mathcal{S}-1$ new sites. (iv) The walker crosses the left desert and
reaches the origin after $\mathcal{S}$ steps.  For the maximal value
$\mathcal{R}^{\dagger}$ of the renewal time, the walker finally reaches the
origin at the end of stage (iv) at the time step where the origin
regenerates.

We now determine $\mathcal{R}^{\dagger}$ by counting the number of steps on
this trajectory. The walker eats the site on the right of the origin at time
step 1, and then takes a time $\mathcal{S}(\mathcal{S}-2)$ if $\mathcal{S}$
is odd, and $(\mathcal{S}-1)(\mathcal{S}-2)$ if $\mathcal{S}$ is even to
complete the phase (i). Similarly, the phase (iii) lasts
$\mathcal{S}(\mathcal{S}-1)$ if $\mathcal{S}$ is odd and $(\mathcal{S}-1)^2$
if $\mathcal{S}$ is even. Moreover, the phases (ii) and (iv) both last
$\mathcal{S}$ steps independent of the parity of $\mathcal{S}$.  Assembling
these results yields the critical value:
\begin{equation}
\mathcal{R}^{\dagger}= 
\begin{cases}
  2 \mathcal{S}^2-3\mathcal{S}+4 \qquad \textrm{$\mathcal{S}$ even}, \\
  2 \mathcal{S}^2-\mathcal{S}+1 \qquad~\; \textrm{$\mathcal{S}$ odd}. 
\end{cases}
\end{equation}

\end{document}